\documentclass[conference]{IEEEtran}

\usepackage{amssymb}           
\usepackage{graphics,graphicx,amssymb,amsmath,amsfonts,verbatim}
\usepackage{epsfig}
\usepackage{subfigure}
\usepackage[subnum]{cases}
\usepackage{color}
\usepackage{lineno}
\usepackage{epstopdf}
\ifCLASSINFOpdf
\else
\fi
\begin{document}
\title{A Zynq-–based flexible ADC architecture combining real-time data  streaming and transient recording}

% \author{\IEEEauthorblockN{Michael Shell}
% \IEEEauthorblockA{School of Electrical and\\Computer Engineering\\
% Georgia Institute of Technology\\
% Atlanta, Georgia 30332--0250\\
% Email: http://www.michaelshell.org/contact.html}}
\author{\IEEEauthorblockN{A.~Rigoni, G.~Manduchi, M.~Gottado, R.~Cavazzana, M.~Recchia, C.~Taliercio, A.~Luchetta}
\IEEEauthorblockA{Consorzio RFX, Corso Stati Uniti 4, Padova 35127, Italy}}


%Telephone: (800) 555--1212, Fax: (888) 555--1212}


\maketitle

% \begin{abstract}
% The abstract goes here.
% \end{abstract}
\begin{abstract}
The RFX-mod2 Nuclear Fusion experiment is an upgrade of RFX-mod, shutdown in 2016. Among the other improvements in the machine structure and diagnostics, a larger number of electromagnetic probes (EMs) is foreseen to provide more information about plasma instabilities and to allow an improved real-time plasma control. An Analog to Digital Converter (ADC) architecture able to provide at the same time both transient recording and real-time streaming, as well as FPGA-based time integration of the inputs is foreseen in RFX-mod2. Transient recording provides full speed data acquisition (up to 1 MSample/s) by recording data in local memory and reading memory content after the plasma discharge. Real-time streaming of sub-sampled data is required for active control. The chosen technology is based on the XILINX Zynq architecture that provides in the same chip a multi-core ARM processor tightly coupled to a FPGA. 
~
Time critical functions are carried out by the FPGA, such as the management of the circular data buffer, low pass filtering for subsampling of the samples to be streamed and digital integration. Other functions are carried out by the processor, such as the management of the configuration setting, received via TCP/IP or HTTP, the data readout of acquired samples in transient recording buffers and network data streaming of data collected for active real-time plasma control. 
\end{abstract}


% \fi
\IEEEpeerreviewmaketitle

\section{Introduction}\label{section:Intro}
~
RFX-mod~\cite{SONATO2003161} is a medium size toroidal plasma multi-configuration machine (major radius  $R=2.0 m$, minor radius $a=0.46m$, operated up to 2 MA current reversed field pinch (RFP) configuration or 0.5 T tokamak). The experiment was shut down in 2016 and is now being upgraded as RFX-mod2~\cite{Peruzzo2018}. A major  foreseen development is a substantial improvement of the magnetic measurement system. In the new experiment configuration many of the EM sensors will be moved inside the vacuum vessel widening their usable signal bandwidth up to 200~kHz in order to providing better plasma control. Moreover, the total number of the new magnetic pick-up coil sensors will be increased with the aim of improved spatial resolution~\cite{MARCHIORI2017892}.

In order to collect magnetic field measurements from EM probes, analog integration system~\cite{pomaro2005transducers} was implemented in RFX-mod, followed by two separate sets of ADC channels, one for precision off-line transient data and the other for real-time control. Re-implementing the same front-end for an increased number of channels is costly, requiring enhanced analog integration and duplication in ADC channels.  A more compact and cost effective solution is being investigated~\cite{gottardo18}, using a configurable FPGA to handle ADC conversion and providing a set of on-line functions directly performed at the FPGA logic level, including the numeric integration in real-time, recording at the same time the dB/dt signals deriving directly from EM coils needed to study the Magneto Hydro Dynamic (MHD) processes taking place into the plasma~\cite{zuin2009current}~\cite{innocente2014tearing}. The possibility of directly acquiring the time derivative of the electromagnetic fields, i.e. the direct signals from EM probes, was not present in the previous system, acquiring integrated signals in order to reduce the number of required ADC channels. This fact introduced a severe limitation in the derivative control required for MHD stabilization because of the bad quality of the computed time derivative.

The proposed approach will further reduce the number of ADC channels by merging high frequency transient recording in local memory (up to 1 MHz) and lower frequency streaming (up to 10 kHz) required for real-time plasma control and having a single ADC channel performing both. In RFX-mod a fixed subset of signals from EM probes was used for the active plasma control, requiring a new set of ADC converters in respect to the transient recorders used for data acquisition. In RFX-mod2 it will be possible to re-use any ADC channel from EM probes for real-time plasma control, being the actual number possibly limited by other factors not related to the ADC devices, such as network bandwidth or control computation load. 

The flexibility provided by a configurable on board FPGA allows also the inclusion of more sophisticated triggering mechanisms and a deeper integration with the timing systems. Examples of triggering mechanism are given by the acquisition of fast transients requiring high speed sampling only in a given, dynamic Region of Interest (ROI). This feature has been implemented in the first proof-of-concept device described in a later section. Deeper integration with the timing systems imply the ability of getting the clock and the trigger signals not only from digital inputs, but also from the specifically coded signal carrying both clock and trigger information (timing highway)\cite{dio4}. Such signals were used in the RFX-mod timing systems to distribute a synchronous clock and asynchronous triggers and a timing device was required for every ADC rack to extract the clock and the trigger signals. The timing device is no more required for a rack hosting the new ADC devices because the ADC devices can directly extract timing information from the timing highway.  

The adoption of a System on Chip (SoC) based technology exploiting both an ARM based processing unit and a FPGA logic provides the  flexibility of a configurable device for real-time operations and as well as the possibility of deploying software components directly on-board. The Red Pitaya board~\cite{redpitaya} is currently used for the development of the architecture. A different solution is however foreseen for the production system integrating an external ADC section with the Zynq-based SoC board. An ADC front end, already used in other applications of real-time plasma control~\cite{ATCA-MIMO-ISOL} was initially considered, but its noise characteristics, and in particular the noise dependency on frequency, proved to limit the quality of digital integration. For this reason, a different solution for the ADC stage is being considered.

The paper is organized as follows: section 2 introduces the reasons for the choice of the Zynq SoC and summarizes the development process. The complete development is orchestrated by the ANACLETO~\cite{rigoni2018framework} framework providing a seamless integration of the different components (VIVADO XILINX tool, compilers, makefiles, ...) required for both FPGA HDL programming and GNU Linux driver development. \\
Section 3 presents the implementation of the ADC flexible architecture providing support for signal integration, subsampling for real-time streaming, ROI detection and timing extraction. 
Section 4 presents the required ADC noise characteristics and discusses the limitations found in existing implementation, along with possible solutions.
Section 5 presents a proof-of-concept system developed under ANACLETO on the Red Pitaya board and providing streamed event-driven high speed data acquisition. Even if not covering all the presented features, this system has been successfully adopted in the NIO negative ion beam experiment~\cite{DEMURI2015249}, a satellite experiment of RFX for the study of additional heating in fusion devices.
%%
%% New ADC DEVICE
%%
\section{The ANACLETO framework for SoC development}
~
As stated before, the RFX-mod2 improved controllability will require the design of a new ADC architecture able to provide both transient recording and real-time streaming. The transient recorder functionality will provide full speed data acquisition (up to $1\approx2$ MSample/s) by recording data in local memory and reading back the memory content after the plasma discharge. At the same time the Real-time streaming is required when the target signal is used in active control because such data must be promptly available in the closed loop feedback. In this case a 10 kHz sub-sampled version is streamed out toward the control units. The solution proposed comes form the adoption of the technology based on the XILINX Zynq architecture that provides in the same chip a multi-core ARM processor tightly coupled to an FPGA. The combined usage of FPGA and a processing unit (running GNU Linux) allows partitioning of the system functionalities into time-critical components mapped onto the FPGA, letting the processor address less critical and possibly more complex functions. 
Developing critical and non critical functions requires HDL development for the first ones, and C or C++ code development for the others. In addition, a GNU Linux driver must be written, acting as a bridge between the FPGA functions and the outside world. Several tools are available from XILINX, including the VIVADO framework for the development and integration of the HDL code into the Zynq architecture, the toolchain required to compile the driver and the support code in the Zynq processor (dual core ARM) and the linux code for that processor. Even if the various components are available, getting them from the proper sources over the network, installing the tools and building the toolchain is a complicated and error-prone process. The ANACLETO framework provides an effective solution by transparently orchestrating the download and the installation of the required components and toolchain, letting the developer concentrate on the specific aspects of his/her project. The project specific tasks, i.e. the development of the HDL code for the FPGA components and their assembly into the firmware project cannot of course be carried out by the ANACLETO, but the framework provides useful hints for the development of the linux driver. This is achieved by ANACLETO by recognizing in the FPGA project what are the components used for the communication with the processor, and producing generic driver templates implementing the communication with the I/O components defined in the FPGA project. Any combination of the following is supported by ANACLETO:
\begin{itemize}
\item I/O registers
\item Input and/or output FIFOs
\item Input and/or output DMA
\end{itemize}
ANACLETO recognizes the I/O registers, FIFOs and DMA controllers defined in the FPGA project (available as IPs in the XILINX toolbox), modifies  the device tree produced by VIVADO and produces the corresponding source code driver template. Of course, no specific functionality can be provided in the driver templates, except for Input and Output data flow management. Starting from the template, the developer will implement the specific functions, but he can ignore to a large extent the intricacies required for the I/O data transfer such as Interrupt handler programming and DMA engine configuration.   
%%
%% Implementation using cheap SoC board
%%
\section{Flexible ADC SoC implementation}
~
The first implementation of the flexible ADC architecture has been carried out on a Red Pitaya board, using the in-board ADC channels. Even if not intended to represent the final application, development of FPGA logic on Red Pitaya offers the advantage of a ready-to-use ADC channel for development and first tests. Most of the firmware will be retained in the final implementation, using a different ADC front end. 

The time critical functions carried out by the FPGA in this context are:
\begin{itemize}
\item The management of a circular data buffer and the DMA transfer in RAM of pre and post trigger samples after the trigger has been received;
\item anti-aliasing filtering and subsequent sub-sampling of the samples to be streamed. The resulting samples are enqueued in a FIFO accessed by the processor;
\item digital integration for deriving magnetic field measurements from EM probe signals. Observe that in this case a single ADC stage will generate two ADC channels;
\item ROI detection in case ADC triggers are derived from the signal itself (e.g. over a given signal level threshold);
\item Clock and trigger extraction in case a highway signal is provided by the timing system, encoding both clock an triggers.
\end{itemize}

The less critical functions that will be carried out by the processor unit are:
\begin{itemize}
\item The management of the configuration setting, received via TCP/IP or HTTP. The processor validates the configuration and write the appreciate registers in the FPGA;
\item off-line data readout of acquired samples in transient recording and communication via TCP/IP with the central data acquisition system;
\item network data streaming of sub-sampled data read from the FIFO and sent in UDP packets to the active plasma control system. 
\end{itemize}
In addition to pre-configured blocks from the XILINX toolbox for data buffering, DMA engine, I/O FIFO and registers, three blocks implemented in VHDL carry out the underlying logic. The first block provides the management of clock and triggers that may be either directly derived from digital inputs or rebuild by properly decoding the timing highway input signal. The second block provides programmable input signal elaboration such as low pass filtering for subsampling and integration. The third block will handle the triggering logic and the circular buffer holding pre and post trigger samples. In particular, the trigger may be derived from external signals (via the first block) or derived from the input signal (e.g. when the input level is greater than a given threshold).    

Communication of subsamples streamed data for real-time plasma control is achieved using the XILINX AXI Stream FIFO. The Xilinx AXI Stream FIFO is a Xilinx free software IP that implement a read/write FIFO queue with a well defined communication protocol. A proper connected IRQ line is used to trigger events to the processing unit together with the related set of status and enable registers. In this way data samples are readily available to the linux processor and will be sent using low latency UDP communication to computing nodes carrying out distributer for real-time plasma control. Communication of the data acquired at high speed in the ROI is carried out by a DMA engine, using circular DMA buffers in order to minimize the number of data copies. In this case data will be sent to the central Data Acquisition system via TCP/IP as soon as a ROI has been acquired. Fig. 1 shows the main blocks of the ADC device: the external ADC circuitry, communicating with the FPGA via a serial LVDS link; the FPGA logic, communicating with the processor via registers, FIFO and DMA; the processor components, in kernel and user space. 
~
~
\begin{figure}[ht]
\centering
\includegraphics[width=0.5\textwidth]{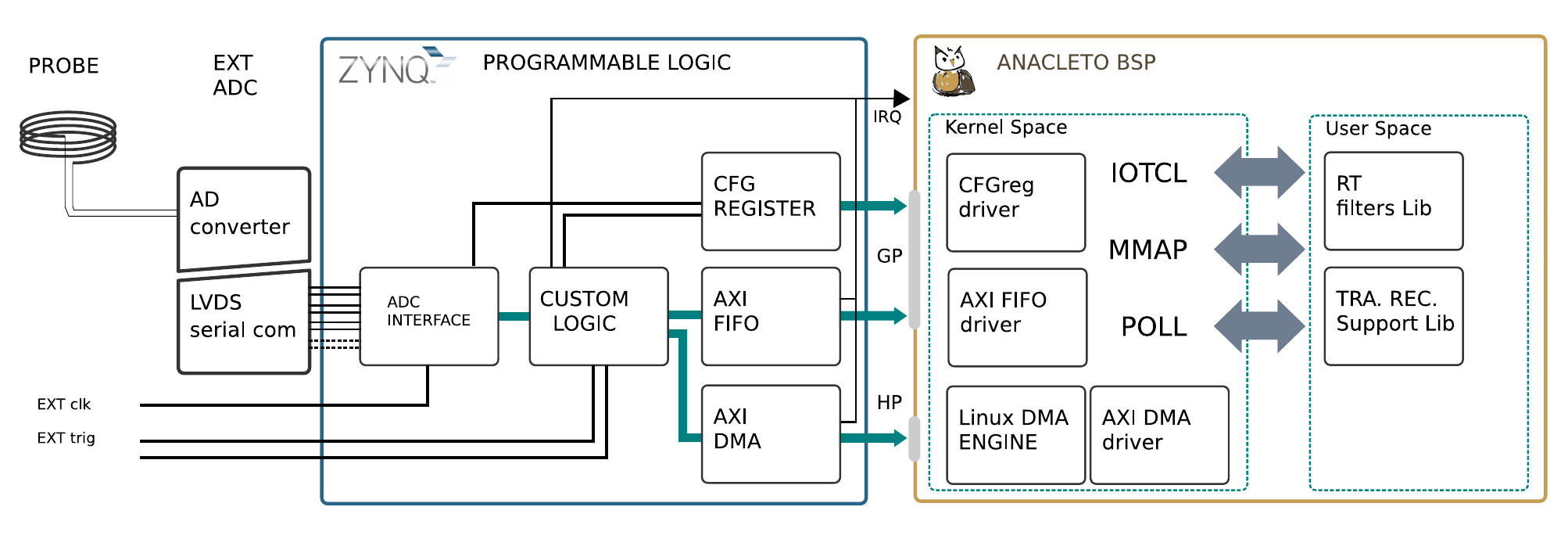}
\caption{Logic design of the flexible ADC structure.}
\label{fig:logic}
\end{figure}
~
% \begin{figure}[ht]
% \centering
% \includegraphics[width=0.49\textwidth]{img/null.png}
% \includegraphics[width=0.49\textwidth]{img/null.png}
% \caption{figure example.}
% \label{fig:1}
% \end{figure}

%%
%% DAQ for RFX-mod2 % ANDREA
%%
\section{Noise requirements in ADC stage for digital integration} \label{daq-example}
~
The Red-Pitaya board, hosting two fast ADC inputs, is currently used for the development of the HDL code and the GNU Linux drivers. Even if it represents a flexible and cheap solution for development, Red-Pitaya is not foreseen in the final production stage. In particular, the strict requests in terms of electrical insulation and noise shape, in order to carry out digital integration, are not met by the converters mounted in this device. 
~
A first prototype has been implemented using a compact and cost effective front end and ADC conversion solution based on the ATCA-MIMO-ISOL~\cite{carvalho2010reconfigurable} architecture that was used for the plasma column vertical stabilization at JET tokamak experiment.
The ADC stage is composed by a plug-in component mounting a 18-bit SAR converter from Analog Devices (AD7641) that acquires signals from a fully differential input in the range 2.048 V at a maximum rate of 2~Msps. The analog input is initially filtered by a 1 pole, 100 kHz passive component connected to an input range adapter (THS4520), then the ADC converter (AD7641) is configured to operate using a 4-wires communication protocol and the electrical insulation is applied to the digital serial data output. The module is completely insulated from the FPGA and the power is supplied through a DC/DC converter. A custom logic has been developed in FPGA to implement the used LVDS serial communication protocol. 

The noise introduced by the ADC front end proved not acceptable for performing numerical integration over a period of some ($<$10) seconds, that is the expected requested integration period in RFX-mod2. The reason is due to the introduced 1/f shaped noise, as shown in Fig. 2. We measured also the noise after bypassing the low pass filter and the impedance adapter, also shown in Fig.2, that turned out to be one order of magnitude less. In a test performed under this condition, the quality of reconstruction of an applied magnetic field collected by a EM probe and integrated in FPGA turned out to be acceptable (Fig. 3). A set of candidate components, whose noise characteristics are compatible with the latter noise level,  has been considered  for the anti-aliasing filter, the impedance adapter and the ADC converter and the selected components will be used for the development of of a new ADC front end component.
~
\begin{figure}[ht]
\centering
\includegraphics[width=0.49\textwidth]{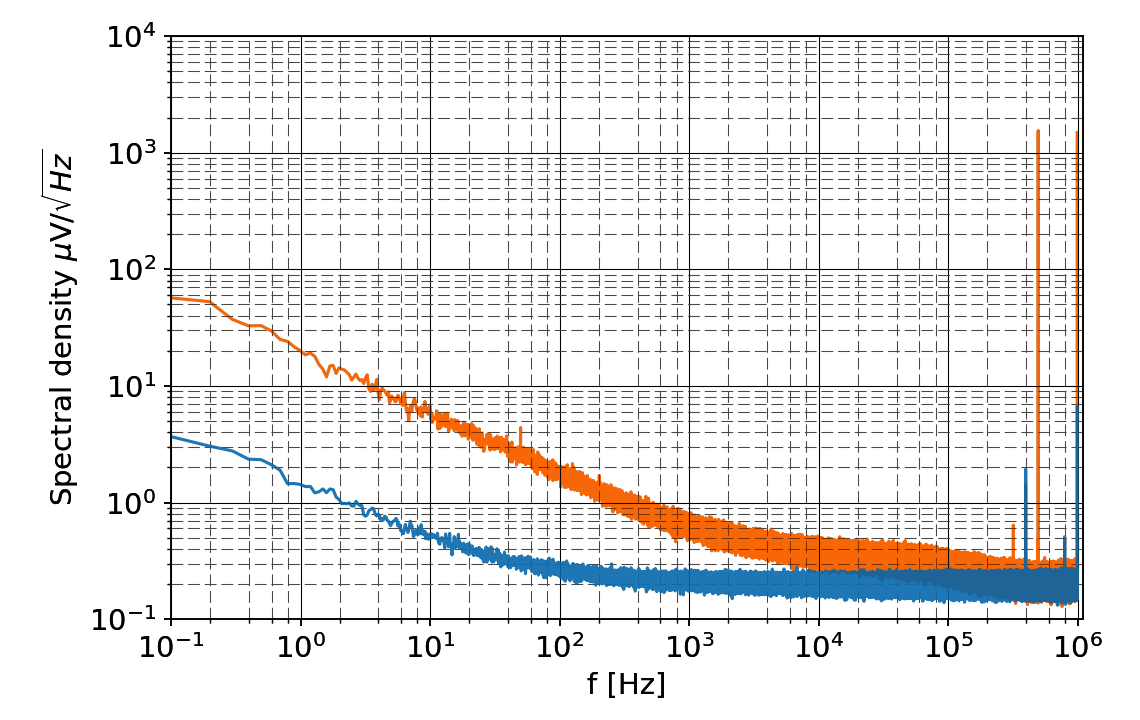}
\caption{Noise spectra.}
\label{fig:noise_spectra}
\end{figure}
~
\begin{figure}[ht]
\centering
\includegraphics[width=0.49\textwidth]{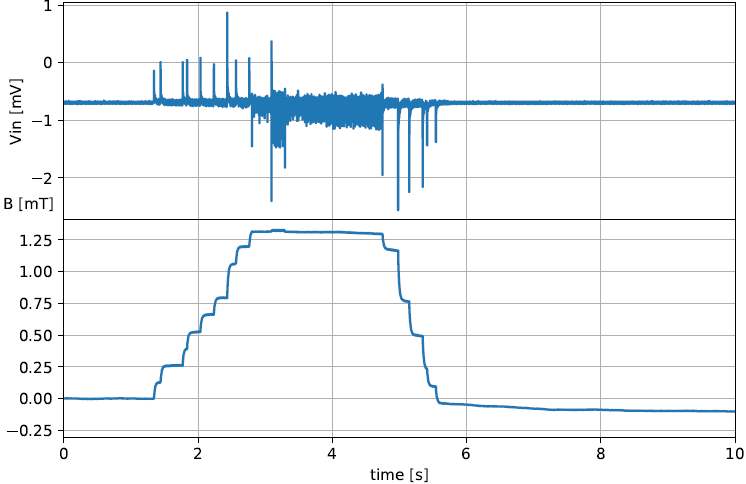}
\caption{test of numerical integrated signal from actual probe with real field applied.}
\label{fig:1}
\end{figure}
~
%%
%% NIO ADC % ANDREA
%%
\section{A proof-of-concept implementation: fast streamed event driven data acquisition for the NIO negative ion beam}
~
Another desired topics for a DAQ device is the possibility to increase the level of detail during acquisition based on particular events. It is not uncommon to have an observed quantity that changes rapidly in time and than last steady or possibly in a non interesting state for long periods. An example of is the breakdown event that occurs in the accelerator grids of a ion source, leading to a very fast transient change in the measured currents and voltages of the grid power supply. In this case, fast data acquisition must be triggered by the event itself, acquiring data for a short time window around the event occurrence. This technique has been applied to Nio experiment~\cite{DEMURI2015249} a small radio frequency negative ions beam source with a high voltage electrostatic particle accelerator stage composed of grids. In certain conditions break-down events~\cite{RECCHIA20111545} appear on the high voltage gaps of the grids causing a  high current discharges of the power supply feeding the accelerator. 
~
A subset of the FPGA functionality described in section 3 has been implemented in a Red Pitaya device, namely the trigger logic to detect the occurrence of the event, the pre and post trigger sampling logic and the FIFO/DMA data transfer to computer memory via the GNU Linux driver. In this case data are streamed and when an event is detected and data collected at 5 MHz sampling speed along the corresponding time window, the data block is passed, either via FIFO or DMA, to the linux driver and then, in turn, to a program in the Zynq processor that communicates the newly acquired data block to the central data acquisition system via TCP/IP. The results are displayed in Fig. 4, showing the events acquired during a beam generation lasting 2 hours. Each time window lasts 1 ms, and one enlarged event is  displayed in the lower part of Fig. 4.    

\begin{figure}[ht]
\centering
\includegraphics[width=0.49\textwidth]{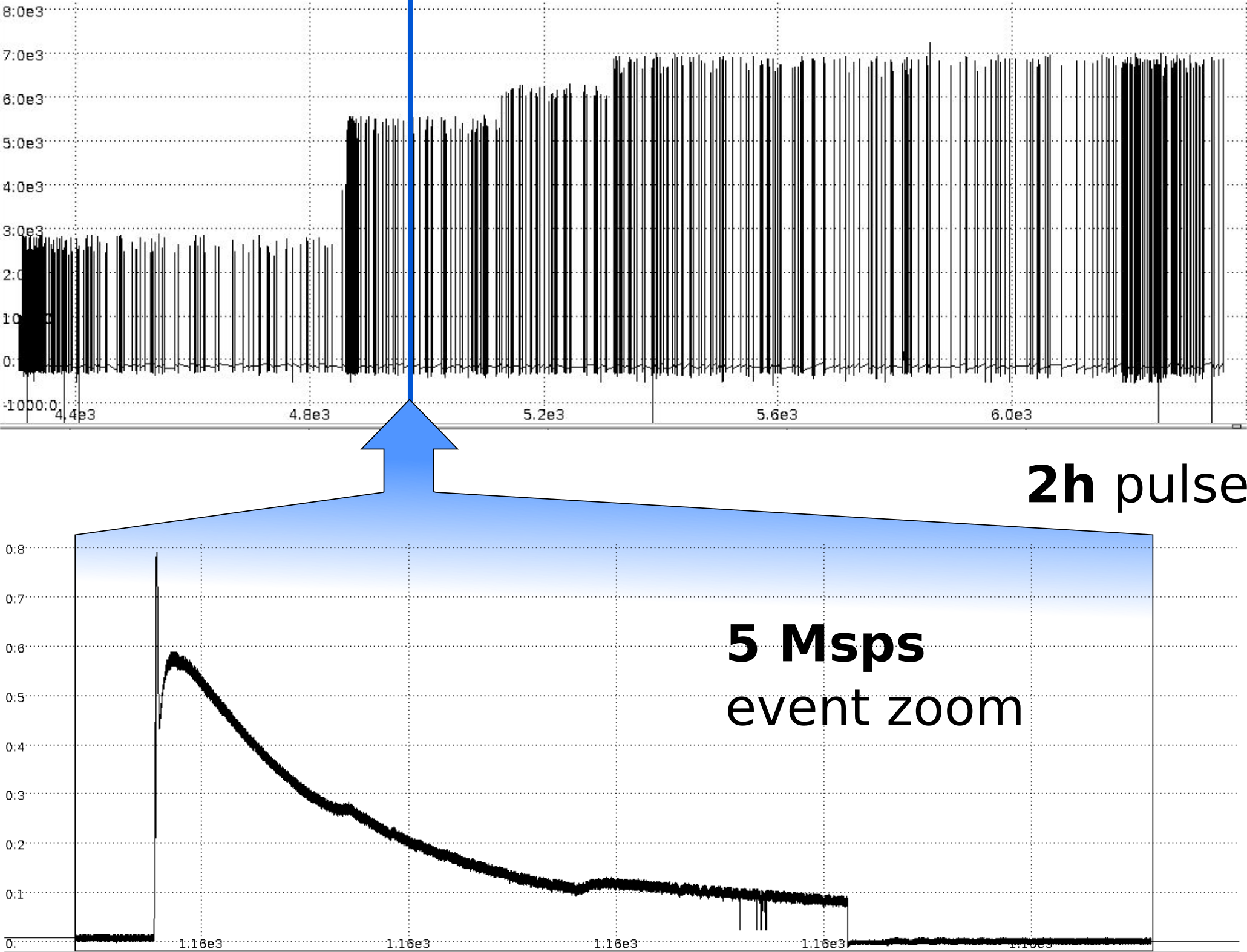}
\caption{figure example.}
\label{fig:nio}
\end{figure}

%
% CONCLUSIONS
% 
\section{Conclusions}\label{section:conclusions}
The architecture of a a new flexible ADC device has been presented, aiming at reducing the number of actual ADC channels by integrating high speed transient recording and data streaming for real-time plasma control. In  addition the same ADC will be used to acquire both magnetic fields and their time derivatives, providing FPGA based digital integration for the derivation of the former. This new architecture is foreseen to be applied at the RFX-mod2 experiment, providing a significant cost reduction in respect to the duplication of devices for transient recording and streaming. Moreover, performing digital integration in place of the analog one before ADC conversion improves the quality of plasma control that can now rely on original time derivative signals. The use of the ANACLETO framework proved to be extremely useful in the development process and shortened quite a bit the learning time for new developers. 

~


% \section*{Acknowledgment}

% The authors would like to thank 



\bibliographystyle{IEEEtran}
\bibliography{REC}


% \end{thebibliography}

\end{document}